\documentclass[12pt,notoc]{JHEP}

\usepackage{amsmath2000,amssymb,euscript,array,cite} 

\input{epsf}
\newcommand{\startappendix}{
\setcounter{section}{0}
\renewcommand{\thesection}{\Alph{section}}}
\newcommand{\Appendix}[1]{
\refstepcounter{section}
\begin{flushleft}
{\large\bf Appendix\thesection: #1}
\end{flushleft}}

\usepackage{epsfig}

\def\N{{\cal N}}

\def\ttau{{\tilde\tau}}

\def\Tr{{\rm Tr}}

\def\det{{\rm det}}

\def\Dbarslash{\,\,{\raise.15ex\hbox{/}\mkern-12mu {\bar\D}}}
\def\Dslash{\,\,{\raise.15ex\hbox{/}\mkern-12mu \D}}
\def\delslash{\,\,{\raise.15ex\hbox{/}\mkern-9mu \partial}}
\def\delbarslash{\,\,{\raise.15ex\hbox{/}\mkern-9mu {\bar\partial}}}

\def\VEV#1{\left\langle #1\right\rangle}

\def\pa{\partial}

\newcommand{\be}{\begin{equation}}
\newcommand{\ee}{\end{equation}}
\newcommand{\bea}{\begin{eqnarray}}
\newcommand{\eea}{\end{eqnarray}}

\newcommand{\EQ}[1]{\begin{equation} #1 \end{equation}}
\newcommand{\AL}[1]{\begin{subequations}\begin{align} #1
\end{align}\end{subequations}}
\newcommand{\SP}[1]{\begin{equation}\begin{split} #1 \end{split}\end{equation}}

\title{Exact Superpotentials from Matrix Models} 
\author{Nick Dorey, Timothy J.~Hollowood, S.~Prem~Kumar and Annamaria
Sinkovics\\
Department of Physics, University of Wales Swansea,
Swansea, SA2 8PP, UK\\
E-mail: {\tt t.hollowood@swan.ac.uk}, {\tt n.dorey@swan.ac.uk}
{\tt s.p.kumar@swan.ac.uk}, {\tt a.sinkovics@swan.ac.uk}}
\preprint{ SWAT-350}
\abstract{
Dijkgraaf and Vafa (DV) have conjectured that the exact superpotential for a
large class of ${\cal N}=1$ SUSY gauge theories can be extracted from
the planar limit of a certain holomorphic matrix integral. We test their
proposal against existing knowledge for a family of deformations of
${\cal N}=4$ SUSY Yang-Mills theory involving an arbitrary polynomial
superpotential for one of the three adjoint chiral superfields.
Specifically, we compare the DV prediction for these models with earlier
results based on the connection between SUSY gauge theories and
integrable systems. We find complete agreement between the two
approaches. In particular we show how the DV proposal allows the
extraction of the exact eigenvalues of the adjoint scalar in the
confining vacuum and hence computes all related condensates of the
finite-$N$ gauge theory. We extend these results to include  
Leigh-Strassler deformations of the $\N=4$ theory.
}

\begin{document}
\setcounter{equation}{0}
\section{Introduction}

Theories with ${\cal N}=1$ supersymmetry in four dimensions are of
considerable interest. 
Apart from their potential phenomenological relevance, they provide a 
field theoretic laboratory where physical phenomena
such as confinement and dynamical symmetry breaking can be studied in a
controlled setting. This is possible largely because these
theories contain a special class of observables whose dependence on the
parameters of the theory is constrained to be holomorphic. These quantities, 
which include the vacuum expectation values of chiral operators and
the tensions of BPS saturated domain walls, are governed by a
holomorphic superpotential which can sometimes be determined
exactly. Although exact results for many different models have been
obtained, a general recipe for computing the
superpotential for an arbitrary ${\cal N}=1$ model has remained
elusive. In recent work \cite{Dijkgraaf:2002vw, Dijkgraaf:2002fc, talk, Dijkgraaf:2002dh}, 
Dijkgraaf and Vafa (DV) have proposed a solution
to this problem which applies to a large class of ${\cal N}=1$
theories. Specifically, they claim that the superpotentials for these
theories can be extracted from the planar limit of a certain
holomorphic matrix integral. In this paper, we will perform a detailed
test of their proposal against known results for a family of ${\cal
N}=1$ preserving deformations of ${\cal
N}=4$ SUSY Yang-Mills theory with gauge group $SU(N)$ (or $U(N)$). 
For this family of theories there is an alternative method for
computing the exact superpotential, given in \cite{nick,nickprem,oferandus,DS}. In fact, the
superpotential can be deduced from the value of the Lax matrix of a
certain classical integrable system in its equilibrium
position. Remarkably, we will find that these two seemingly very
different recipes give exactly the same result. The comparison will
also shed light on both approaches. In the remainder of this introductory
section, we will review both the DV proposal and the results of
\cite{DS} based on the connection to integrable systems 
and present our main conclusions. 

We begin by considering ${\cal N}=4$ SUSY Yang-Mills theory with gauge
group $G=SU(N)$ and bare complex coupling $\tau=4\pi i/g^{2}_{YM} +
\theta/2\pi$. In the standard language of ${\cal N}=1$ superfields,
the theory contains three adjoint-valued chiral multiplets, which we
will label $\Phi^{+}$, $\Phi^{-}$ and $\Phi$. We will consider the
deformation of the ${\cal N}=4$ theory obtained by adding an ${\cal N}=2$
preserving mass term for $\Phi^{\pm}$, together with a general
polynomial superpotential for $\Phi$ which preserves ${\cal N}=1$ 
supersymmetry.The classical superpotential then reads, 
\begin{equation}
W(\Phi)= \Tr\;\left(i\Phi[\Phi^{+},\Phi^{-}] \, + \, 
\Phi^{+}\Phi^{-} \, +\, \sum_{p=2}^{N}{g_{p}}\Phi^{p}\right) 
\label{spcl}
\end{equation}
Setting $g_{p}=0$ for $p\geq 3$ corresponds to the ${\cal N}=1^{*}$
deformation but we will not limit ourselves to this case. 
\paragraph{}
This theory already has a rich vacuum structure at the classical
level. In particular, there is a large set of classical vacua in which
the adjoint scalar fields acquire vacuum expectation values (VEVs)
spontaneously breaking some or all of the gauge symmetry. In this
paper we will focus on SUSY vacua in which the full $SU(N)$ gauge
symmetry is unbroken. At the classical level this corresponds to
setting the adjoint scalar VEVs to zero\footnote{In general, the 
adjoint scalar fields can 
and will acquire VEVs in the quantum theory without breaking the gauge
symmetry.}. The qualitative behaviour of the corresponding quantum
theory in these vacua is well known: at energy scales much less than
the mass scale set by the deformations, the chiral multiplets decouple leaving the theory of 
a single $SU(N)$ vector multiplet which is in the same universality
class as ${\cal N}=1$ SUSY Yang-Mills. The theory is therefore in a
confining phase with a mass gap and gluino condensation. 
There are $N$ supersymmetric vacua
denoted $|k\rangle$ with $k=0,1,2,\ldots, N-1$, each with a non-zero 
expectation value for the gluino bilinear which is the lowest component
of the superfield 
$S={\Tr}\;(W_{\alpha}W^{\alpha})/32\pi^{2}$. In fact, the
action of S-duality, inherited from the underlying ${\cal N}=4$
theory, dictates 
that the vacuum states $|k\rangle$ and $|k+1\rangle$ are simply related 
by a $2\pi$ shift in the vacuum angle: $\tau\rightarrow
\tau+1$. Hence it will suffice to consider the $k=0$ vacuum,  
$|0\rangle$, in the following. Our aim will be to determine the exact 
effective superpotential in this vacuum, $\langle 0| W_{\rm eff}|
0\rangle$. The vacuum expectation values of the operators $u_{p}={\Tr}\;\Phi^{p}$ can then be obtained by differentiating with respect
to the parameters $g_{p}$ appearing in (\ref{spcl}).    
\subsection{The integrable system approach}
\paragraph{}
An exact formula for the effective superpotential can be obtained from 
a slight extension of the results presented in \cite{DS}.  
This approach relies on the connection between SUSY gauge theories and 
integrable systems first discovered in \cite{DW,R,MW}. 
In the present case the relevant
integrable system is the elliptic Calogero-Moser (ECM) 
system. This is a system of $N$
non-relativistic particles with complex positions and
momenta, $X_{j}$ and $P_{j}$, 
$j=1,2,\ldots, N$, which live on a torus $E_\tau$ of
complex structure $\tau$ and its tangent space respectively. 
Time evolution is governed by the quadratic Hamiltonian, 
\begin{equation}
H_{2}= \sum_{j=1}^{N} \frac{P_{j}^{2}}{2} + \sum_{j>k} {\wp}(X_{j}-X_{k})
\label{ecm}
\end{equation} 
where ${\wp}$ is the Weierstrass function (see Appendix A.1) which is
elliptic with respect to the torus $E_\tau$. 
The system has a Lax formulation  involving two $N\times N$ matrices
$L_{jk}$ and $M_{jk}$, in terms of which Hamilton's equations can be
rewritten as $i\dot{L}=[M,L]$. This immediately implies the existence
of $N$ conserved quantities $H_{p}={\Tr}\;L^{p}$ for
$p=1,\ldots,N$. These 
quantities can also be thought of as a set of Poisson-commuting 
Hamiltonians each generating a distinct time-evolution of the system.    
\paragraph{} 
We will now state the relation between the deformed ${\cal N}=4$ theory and the
ECM theory (for more details see \cite{DS}). Firstly, for each value of $p$, 
the conserved quantity $H_{p}$ is
identified with the expectation value of the operator $u_{p}={\Tr}\;\Phi^{p}$
introduced above\footnote{To restrict to the $SU(N)$ theory where $u_{1}$
vanishes we consider the ECM system in its center-of-momentum frame
where $H_{1}=\sum_{j=1}^{N}P_{j}=0$.}. 
As explained in \cite{nick}, the supersymmetric vacua
of the theory then correspond to the equilibrium configurations of the 
ECM particles. In particular, for the theory with an arbitrary
polynomial deformation, as in Eq.(\ref{spcl}) we must consider
equilibrium configurations with respect to the time-evolution 
generated by the Hamiltonian $H=\sum_{p=2}^{N}g_{p}H_{p}$. 
\paragraph{}
In the general case, finding these
configurations is not straightforward. However, the massive vacua
of the theory have a very special property: they correspond to 
static configurations
which {\em simultaneously} stationarize each of the Hamiltonians
\footnote{In the field theory this is related to the fact that massive
vacua preserving ${\cal N}=1$ supersymmetry correspond to the
maximal degenerations of the Riemann surface which governs the Coulomb
branch of the ${\cal N}=2$ theory obtained by setting all the
deformation parameters to zero. However, it can also be demonstrated
directly using the Lax representation for the equations of
motion of the ECM system. This is shown in Appendix B.} 
$H_{p}$, for $p=2,\ldots,N$. 
In particular, the vacuum $|0\rangle$ introduced above corresponds to
a configuration where the particles are evenly spaced around one cycle
of the torus: $X_{j}=(2\pi i\tau/N)j-i\pi\tau(N-1)/N$; $P_j=0$ for $j=1,\ldots, N$. As each
Hamiltonian is stationary, the same configuration corresponds to a SUSY
vacuum for all values of the deformation parameters. 
The eigenvalues of the  Lax matrix $L_{ab}$ in this equilibrium
configuration were calculated explicitly in \cite{DS}. The result is  
\begin{eqnarray}
\tilde\lambda_{j} &= & - \frac{1}{2}  
\frac{\theta_{3}'(\pi/2-\pi j/N|\tau/N)}{\theta_{3}(\pi/2-\pi
j/N|\tau/N)} \nonumber \\ 
\label{DSresult}
\end{eqnarray}
where $j=1,2,\ldots, N$. 
The vacuum expectation values of the operators, $u_{p}$, can then be
determined via the relation $\langle 0|u_{p}|0\rangle
=\sum_{j=1}^{N}\tilde\lambda_{j}^{p}$. As the resulting formulae
are completely independent of the deformation parameters $g_{p}$, we
find that that the vacuum value of the effective superpotential is 
simply given by, 
\begin{equation}
\langle 0| W_{\rm eff}|0\rangle = \sum_{p=2}^{N}\, g_{p}
\langle 0|u_{p}|0\rangle
=\sum_{p=2}^{N}\,\sum_{j=1}^{N}\,g_{p}\tilde\lambda_{j}^{p}
\label{res2}
\end{equation}
\subsection{The Dijkgraaf-Vafa Approach}
\paragraph{}
Dijkgraaf and Vafa have proposed an exact recipe for calculating the
exact effective superpotential $W_{\rm eff}$ as a function of the
gluino condensate (the expectation value of the superfield $S$). 
Specifically, they relate the superpotential to the following
holomorphic matrix integral: 
\begin{eqnarray} 
{\cal Z} & = & \int\, [d\Phi] [d\Phi^{+}] [d\Phi^{-}] 
\, \exp\left(- \frac{1}{g_{s}}W(\Phi)\right)
\label{partition}
\end{eqnarray}
where $g_{s}$ is a coupling constant. 
The integral is to be understood as a multi-variable contour integral.
As for any adjoint $SU(N)$ theory the free
energy ${\cal F}$ has a 't Hooft expansion where the Feynman diagrams
are organised according to their genus $g\geq 0$. 
\begin{equation}
{\cal F}= \sum_{g\geq 0} g_{s}^{2g-2} {\cal F}_{g}(\lambda)
\label{free}
\end{equation}
where $\lambda=g_{s}N$ is the matrix model 't Hooft coupling. DV
predict that the exact superpotential of the above $\N=1$ perturbation of the $\N=4$ theory
only depends on the leading term in this expansion: ${\cal F}_{0}(\lambda)$. 
This term may be evaluated by taking the 't Hooft limit of the
matrix model, $N\rightarrow \infty$, $g_{s}\rightarrow 0$ with
$\lambda$ held fixed, which isolates the planar Feynman diagrams. 
Fortunately the matrix model can be solved exactly in this limit using
standard saddle-point technology. In the case of a quadratic
deformation, the solution has been already been given by Kazakov,
Kostov and Nekrasov \cite{Kazakov:1998ji}. Generalization of their solution 
to an arbitrary polynomial deformation is straightforward and is given
below. 
\paragraph{}
The proposed dictionary between matrix model and field theory
quantities identifies the 't Hooft coupling $\lambda=g_{s}N$ with the
expectation value of the gluino bilinear superfield $S$. From now on
we will use the symbol $S$ to denote both the matrix model 't Hooft
coupling and the gluino condensate of the field theory and assume that
the interpretation is clear from the context.    
The resulting expression for the field theory superpotential is, 
\begin{equation}
W_{\rm eff}(S)=\, N \frac{\partial {\cal F}_{0}}{\partial S} \, -\,
2\pi i \tau S 
\label{wdv}
\end{equation}
\paragraph{}
The supersymmetric vacua may be found be extremising $W_{\rm eff}$ as
a function of S. The expectation value of $W_{\rm eff}$ in each vacuum
is simply its value at the corresponding the extremum. 
The main aim of this paper is to compare the resulting formulae with
the exact field theory predictions Eqs. \eqref{DSresult} and \eqref{res2}.
We find that the DV prescription allows us to extract the field theory
eigenvalues $\tilde\lambda_j$     
and we find precise agreement between the two approaches, for arbitrary
values of the deformation parameters $g_{p}$, up to the known
ambiguities in the definition of the operators $u_{p}$ \cite{oferandus}. 
Agreement in the case of a quadratic superpotential has previously been
checked in \cite{Dijkgraaf:2002dh}.   

Several features
of our result are worthy of further comment. A remarkable consequence
of Dijkgraaf and Vafa's proposal is that the holomorphic
quantities of the deformed $\N=4$ theory are large-$N$
exact. Although this property is far from obvious in the integrable
systems approach it nevertheless seems to be true that $1/N$ corrections to all
physical quantities vanish up to the usual operator mixing ambiguities!
On another hand there are some features of the field theory results
which are not apparent at the outset in the DV prescription, for 
example the linear dependence of the exact superpotential (\ref{res2}) 
on the deformation parameters $g_{p}$. Nevertheless, we are able to
prove that for the class of deformations considered the DV
prescription automatically implies this linearity in the confining vacuum.

We also extend our study to include a different class of deformations
of the $\N=4$ theory, namely the Leigh-Strassler deformations. We use
the DV proposal to compute the exact effective superpotential for
these deformations and extract the eigenvalues
$\tilde\lambda_j$ in the confining vacuum of the theory. Remarkably,
we find once again that these are independent of the deformation
parameters $g_p$.

\section{Superpotentials and eigenvalue distributions
from matrix models}
According to the Dijkgraaf-Vafa proposal 
\cite{Dijkgraaf:2002fc,Dijkgraaf:2002vw,talk}
superpotentials for a 
large class of $\N=1$ supersymmetric field theories at {\it finite} $N$ 
are computed by corresponding large-$N$ matrix models. Assuming the
validity of this conjecture for arbitrary deformations of the $\N=4$
theory, we will demonstrate that for such deformations, in the
confining vacuum of the
resulting theory the matrix model in fact yields the correct
eigenvalue distribution for the adjoint scalar field in the
theory. This tests the Dijkgraaf-Vafa proposal for a large class of  
deformations of the $\N=4$ theory. Furthermore, it 
provides a general prescription for extracting QFT
eigenvalue distributions from the matrix model and also demonstrates
that the matrix model computes all condensates in the holomorphic
sector of the theory.

\subsection{Deformations of $\N=4$ theory}

In the $\N=1$ language, the $\N=4$ $U(N)$ theory with coupling
constant $\tau\equiv 4\pi i/g^2_{YM}+\theta/2\pi$ has three adjoint
chiral superfields $\Phi^+,\Phi^-$ and $\Phi$. We consider a general
class of deformations of the $\N=4$ theory specified by a tree level
superpotential
\EQ{W={1\over
g^2_{YM}}\Tr\left[i\Phi[\Phi^+,\Phi^-]+\Phi^+\Phi^-+V(\Phi)\right]}
where $V(\Phi)$ is a general polynomial
\EQ{V(\Phi)\equiv \sum_p {g_p}\,\Phi^p.}
Note that for the sake of simplicity we have set any and all mass
scales to unity. (For $V(\Phi)=\Phi^2$ we obtain the $\N=1^*$
theory.) 

As a direct consequence of the proposal of Dijkgraaf and Vafa
\cite{Dijkgraaf:2002fc,Dijkgraaf:2002vw,talk}, in a
given vacuum the exact effective 
superpotential for the above class of deformations of the $\N=4$ theory
is computed by the planar diagram expansion, $i.e.$ a large-$N$ limit,
of the three-matrix model partition function expanded around that vacuum
\EQ{Z=\int [d\Phi^+][d\Phi^-][d\Phi]\exp-{1\over
g_s}\Tr\left[i\Phi[\Phi^+,\Phi^-]+\Phi^+\Phi^-+V(\Phi)\right]
\label{mm}} 
with $\Phi^+=(\Phi^-)^\dagger$.

The confining vacua of the field theory correspond to the classical
solution $\Phi=\Phi^\pm=0$ which preserves the full $U(N)$ gauge
symmetry. In order to implement the Dijkgraaf-Vafa proposal for
extracting the field theory superpotential in the confining vacua we
therefore need to solve the matrix
model Eq.~\eqref{mm} around the trivial vacuum $\Phi=\Phi^\pm=0$. As
noted for example in \cite{Kazakov:1998ji}, the key fact that permits
the solution of these matrix models is that one may integrate out
$\Phi^\pm$ exactly to obtain a one-matrix integral to be solved in  the
large-$N$ limit
\EQ{Z=\int[d\Phi] {e^{-{1\over g_s}\Tr V(\Phi)}\over{\det(
\text{Adj}_\Phi+i)}}.}

\subsection{Solution of the large-$N$ matrix model}

The above one-matrix model actually becomes tractable in the large-$N$
limit by going to an eigenvalue basis and performing a large-$N$
saddle-point approximation to the integral. The details of this
procedure have been extensively discussed in the literature. We refer
the reader to \cite{Kazakov:1998ji}, and
references therein, for details. 

The main feature of the solution is that the eigenvalues $\lambda_j$ of
$\Phi$ interact via a repulsive effective potential and form a
continuum in the large-$N$ limit and   
condense into a cut along the real axis. The actual extent of the
cut and the density of eigenvalues $\rho(\lambda)$ along the cut is
self-consistently determined by the saddle point equation in terms of
the parameters of the deformation $V(\Phi)$ and the matrix model 't Hooft
coupling $S=g_sN$. The saddle point equation in the large-$N$ limit is
most conveniently written in terms of the resolvent function
\EQ{\omega(z)=\int_{-\alpha}^{+\alpha}{\rho(\lambda)\over{z-\lambda}}d\lambda
\qquad\lambda\in[-\alpha,\alpha];\qquad\int_{-\alpha}^{+\alpha}\rho(\lambda)d\lambda=1.}
The resolvent $\omega(z)$ is an analytic function on the complex
$z$-plane whose only singularity is a branch cut with the
discontinuity across the cut given precisely by the spectral density
\EQ{\omega(\lambda+i\epsilon)-\omega(\lambda-i\epsilon)=-2\pi
i\rho(\lambda);
\qquad\lambda\in[-\alpha,\alpha].}

The saddle point equation essentially encodes the condition that the force
on a test eigenvalue due to the remaining large-$N$
equilibrium distribution of eigenvalues vanishes along the cut where
the eigenvalues condense. The matrix model effective action yields
this force on a test eigenvalue at position $z$ in the complex plane as
\EQ{F(z)= {1\over 2i}[G(z+\tfrac i2)-G(z-\tfrac i2)]}
where
\EQ{G(z)=U(z)+iS(\omega(z+\tfrac i2)-\omega(z-\tfrac i2)).
\label{defG}}
and $U(z)$ is a polynomial in $z$ such that
\EQ{V^\prime(z)={1\over i}[U(z+\tfrac i2)-U(z-\tfrac i2)].\label{defU}}

The analytic properties of the function $G(z)$ are central to the
solution of the model and its subsequent physical
interpretation. The defining equation for $G(z)$ Eq.~(\ref{defG})
implies that $G(z)$ is an analytic function with two branch
cuts between $[-\alpha+{i\over 2}, \alpha+{i\over 2}]$ and
$[-\alpha-{i\over 2},\alpha-{i\over 2}]$. The matrix model saddle
point equation identifies the values of $G(z)$ along the two cuts via
\EQ{G(\lambda+\tfrac i2\pm i\epsilon)=G(\lambda-\tfrac i2\mp
i\epsilon);\qquad\lambda\in[-\alpha,\alpha].\label{glu}}
Thus the saddle point equation may be viewed as a `gluing' condition
for the two cuts on the $z$-plane. In addition $G(z)=G(z^*)^*$ and
$G(z)=G(-z)$. The function $G(z)$ is then uniquely determined by its
asymptotic behaviour at $z\rightarrow \infty$ which is specified by
the polynomial $U(z)$ in Eq.~(\ref{defG}).

Given the density of eigenvalues $\rho(\lambda)$ in the matrix model
one may use it to compute expectation values of various quantities in
the matrix model. However, at the outset we would like to point out
these expectation values are distinct from vacuum expectation values
of the corresponding observables in field theory. In particular,
expectation values of the quantities ${\rm Tr}\,\Phi^n$ in the matrix
model are defined in the obvious way:
\EQ{
\langle\langle{\rm Tr}\,\Phi^n\rangle\rangle=\int_{-\alpha}^\alpha
\rho(\lambda)\lambda^n\,d\lambda\ .
\label{exp}
}
Using the formulae above one can relate these expectation values to the
terms in the large-$z$ expansion of $G(z)$. However these expectation values
$\langle\langle{\rm Tr}\,\Phi^n\rangle\rangle$ are {\it not\/} to be
identified with the corresponding condensates $\VEV{\Tr\Phi^n}$in the
field theory,
\EQ{\VEV{\Tr\,\Phi^n}_{QFT}\neq\langle\langle{\rm Tr}\,\Phi^n\rangle\rangle. }

\subsection{Emergence of the $\ttau$-torus}

The two-cut structure of the $z$-plane (from the analytic
properties of $G(z)$) and the gluing conditions Eq.~\eqref{glu} uniquely
specify a torus which we denote by $E_\ttau$ where $\ttau$ is defined
to be the complex
structure parameter of this torus. 
The contour ${\cal C}_A$ in the $z$-plane enclosing the cut
$[-\alpha+{i\over 2},\alpha+{i\over 2}]$ maps to the $A$-cycle of the
torus while the contour ${\cal C}_B$ joining the two cuts maps to the
$B$-cycle of the torus. We also define ${\cal C}_{A'}$ as the contour
enclosing the lower cut. Note that ${\cal C}_A+{\cal C}_{A'}$ can be
pulled away from the cuts to enclose the point at infinity. These
contours in the $z$-plane are indicated in Figure 1.
\begin{figure}
\begin{center}\mbox{\epsfysize=5.5cm\epsfbox{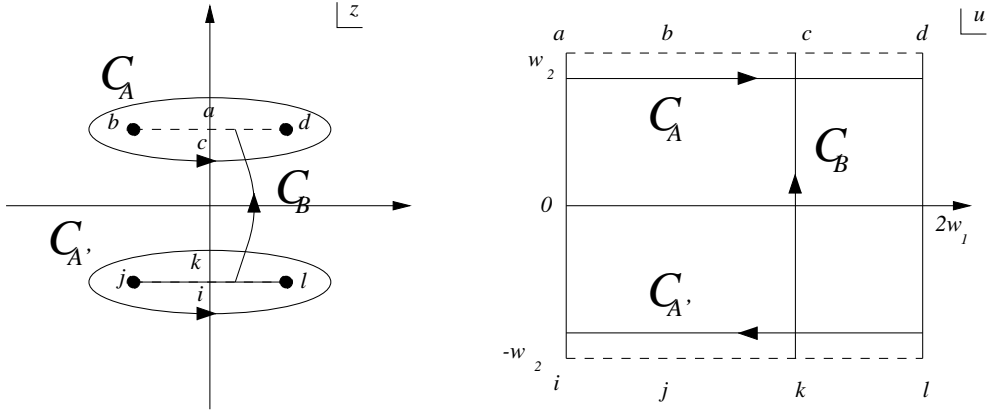}}\end{center}
\caption{\small The $z$-plane and its pre-image in the $u$-plane. The
cuts in the $z$-plane are indicated by the dotted lines with Roman
letters to dictate how they are mapped from the $u$-plane. Three
representative closed contours ${\cal C}_A$, ${\cal C}_B$ and ${\cal
C}_{A'}$ are illustrated.}
\end{figure}

The torus $E_\ttau$ may be thought of as usual, as a quotient of the
complex plane, say the $u$-plane (this auxiliary variable $u$ will lead
to the required elliptic parametrization), by a lattice with
periods $2\omega_1$ and $2\omega_2$ with
$\omega_2/\omega_1=\ttau$. The physical significance of this torus
will become apparent upon implementing the procedure of
\cite{Dijkgraaf:2002fc,Dijkgraaf:2002vw,talk} 
for computing the field theory
superpotential. In fact $\ttau$ will be identified with $\tau/N$,
signalling a low-energy $SL(2,{\mathbb Z})$ duality symmetry in
$\tau/N$ where $\tau$ is the coupling of the $\N=4$ theory. This
duality, dubbed $\tilde S$-duality in \cite{oferandus} is distinct
from the $SL(2,{\mathbb Z})$ symmetry of the $\N=4$ theory although it
is of course a consequence of the latter.

The map $z(u)$ from the $u$-plane to the $z$-plane is specified by the
requirements that going around ${\cal C}_A$ (the $A$-cycle of the
torus) returns
$z$ to its original value, while traversing the contour
${\cal C}_B$ (the $B$-cycle of the torus) causes $z$ to jump by an
amount $i$, which is the distance between the two cuts in the
$z$-plane. Both these operations leave $G$ unchanged which is
therefore an elliptic function on the $u$-plane. Thus 
\AL{&A-{\rm cycle:}\quad z(u+2\omega_1)=z(u);\qquad
G(z(u+2\omega_1))=G(z(u))\ ,\\
&B-{\rm cycle:}\quad z(u+2\omega_2)=z(u)+i;\qquad
G(z(u+2\omega_2))=G(z(u)).
} 

These requirements (along with the power law behaviour of $G(z)$ at
large $z$) specify $z(u)$ uniquely and we find that
\EQ{z(u)=-{\omega_1\over
\pi}\left[\zeta(u)-{\zeta(\omega_1)\over\omega_1 }u\right]
\equiv -{1\over2}{\theta_1^\prime({\pi
u\over2\omega_1}|\ttau)\over\theta_1({\pi u\over2\omega_1}|\ttau)}.\label{defz}}
Here $\zeta(z)$ is the Weierstrass zeta function and $\theta_1$ is the
Jacobi theta function (see \cite{WW} and the Appendix for details).
Note that the map from the torus to the eigenvalue plane is completely
independent of the choice of the field theory deformation
$V(\Phi)$. Different deformations imply different asymptotic
properties of the elliptic function $G(z(u))$ as $z\rightarrow\infty$
and hence lead to elliptic
functions with different orders. 

\subsubsection{An Example}

For the simplest deformation, {\it i.e.\/}~the 
quadratic one corresponding to the $N=1^*$ theory, the asymptotics determine
$G(z(u))$ uniquely:
\EQ{V(\Phi)=\Phi^2:\qquad G(z(u))={\omega_1^2\over \pi^2}
\Big[{\wp}(u)-2{\zeta(\omega_1)\over\omega_1}\Big]\label{gquad}.}
$G(z(u))$ will involve higher powers of the Weierstrass-$\wp$
function for more complicated deformations. 

Let us see how in the simplest example which was first discussed and solved
in \cite{talk, Dijkgraaf:2002dh}, $V(\Phi)=\Phi^2$, $\ttau$ gets identified
with $\tau/N$. To see this we integrate the one-form $G(z)dz$ over the
$A$-cycle and $B$-cycle to obtain the total number of eigenvalues $S$
and $\partial{\cal F}_0/\partial
S$ respectively, the latter being the variation in the genus zero free energy of
the matrix model upon
transporting an eigenvalue to infinity,
\AL{&\Pi_A=2\pi i S=-i\int_AG(z(u)){dz(u)\over du}du={1\over
12}{dE_2(\ttau)\over d\ttau},\label{s}\\
&\Pi_B={\partial{\cal F}_0\over \partial S}=-i\int_B G(z(u)){dz(u)\over
du}du={\ttau\over 12}{dE_2(\ttau)\over d\ttau}-{1\over 12}E_2(\ttau).}
Then the effective superpotential $W_{\rm eff}$ for the $\N=1^*$ theory 
in the
confining vacuum is obtained by extremizing the following expression
with respect to $S$,
\EQ{W_{\rm eff}=N\Pi_B-\tau\Pi_A\ .}
Using our expressions for the respective integrals over the cycles it is
easy to see that,
\EQ{{\partial W_{\rm eff}\over\partial S}=0\ \Longrightarrow\ 
\ttau={\tau\over
N}.}
and 
\EQ{W_{\rm eff}=-{N\over12}E_2({\tau/N})\ .}
Upon restoring the mass scales in the problem this result agrees
precisely, up to an additive constant with the expressions of
\cite{nick,nickprem}. The fact that $S$ in Eq.\eqref{s} automatically
comes out as the 
$\tau$-derivative of the effective superpotential is in line with the
identification of $S$ with gluino condensate of the gauge theory.

The identification of the complex structure $\ttau$ of the matrix
model torus with $\tau/N$ in the confining vacuum of the field theory
in fact holds for all
possible choices of $V(\Phi)$ and indeed, seems to work for a
completely different class of deformations namely the Leigh-Strassler
deformations which will be discussed in detail below. The emergence of
a torus with complex structure $\tau/N$, controlling the low-energy
physics of the confining vacua of
deformations of the $\N=4$ theory, has been well understood from the
point of view of Seiberg-Witten theory in \cite{oferandus,us}. It is
quite remarkable to see this structure emerge from the matrix model.

\subsection{QFT eigenvalues and general superpotentials}

Before embarking on a discussion of the eigenvalues below, 
we should emphasize a subtle point that will emerge in due course. 
It will turn out that the distribution of eigenvalues in the matrix
model described by the density $\rho(\lambda)$ is {\it not\/} 
to be identified
with the distribution of eigenvalues of $\Phi$ in the field
theory at large $N$. In fact, the main result of this section is that
the Dijkgraaf-Vafa proposal is tantamount to
identifying the eigenvalues of the adjoint scalar 
in the confining vacuum of the field theory with the 
function $z(u)$ in Eq.~\eqref{defz} (after a simple
shift of its argument), distributed evenly over a cycle on the 
$E_\ttau$ 
torus. In contrast the matrix model eigenvalues are not uniformly
distributed along this cycle.

The function $z(u)$ that we have obtained above is a map from the
torus $E_\ttau$ to the plane of eigenvalues $z$. In particular, $z(u)$
maps the $A$ cycle of the torus to the cut enclosed by ${\cal
C}_A$,
\EQ{{\cal C}_A\;\simeq\;
z(2\omega_1x+\omega_2)=-{1\over2}{\theta_3^\prime({\pi x}|\ttau)
\over\theta_3({\pi x}|\ttau)}+{i\over 2}; \qquad x\in[-\tfrac12,\tfrac12].}
This cut extends from $-\alpha+{i\over 2}$ to $\alpha +{i\over
2}$ and represents the possible values of the 
eigenvalues $\lambda$ of $\Phi$ shifted by $i/2$
(the $i/2$ shift is deduced simply from Eq.~(\ref{defG})).
In view of this we define the function
\EQ{\lambda(x)=-{1\over2}{\theta_3^\prime({\pi x}|\ttau)
\over\theta_3({\pi x}|\ttau)};\qquad x\in[-\tfrac12,\tfrac12]\ .
\label{eigen}}
At first sight, this is simply a map from the $A$ cycle of the torus to
the interval $[-\alpha,\alpha]$ in which the eigenvalues are
distributed. However, we will soon find that $\lambda(x)$ has greater
significance. Before we uncover this, we note that given the map
$\lambda(x)$ we can ask how the eigenvalues of the matrix model are
distributed along the $A$ cycle of the torus. The answer is rather
complicated being given by  
\EQ{
\frac{dn}{dx}=\rho(\lambda(x))\frac{d\lambda}{dx}\ ,
}
with $-x_0\leq x\leq x_0$ where $\lambda(x_0)=\alpha$.\footnote{Notice
that the $A$ cycle covers the interval $[-\alpha,\alpha]$ twice, so
one must choose one of the branches.}

In contrast, we will now show how the proposal of 
\cite{Dijkgraaf:2002dh}
automatically implies that eigenvalues of $\Phi$ in the confining
vacuum of the field theory
at large $N$ are uniformly distributed over the $x$-interval
$[-\tfrac12,\tfrac12]$ and are given precisely by the function $\lambda(x)$.
To see this we simply note that the function
$z^{\prime\prime}(u)={\omega_1\over\pi}\wp^\prime(u)$ is elliptic on the
torus and thus  the following integral vanishes 

\EQ{\int_{{\cal C}_A+{\cal C}_A^\prime}G(z(u)){z^{\prime\prime}(u)}du=0.}
Using the definition of $G(z)$ in Eq.~(\ref{defG}) this implies
\EQ{S\int_{{\cal C}_A+{\cal C}_A^\prime}(\omega(z+\tfrac i2)-
\omega(z-\tfrac i2))z^{\prime\prime}(u)du=i\int_{{\cal C}_A+{\cal
C}_A^\prime} U(z)z^{\prime\prime}(u)du.}  
The periodicity of $z(u)$ along the $A$-cycle allows us to reduce the
left hand side of this expression to a contour integral around $u=0$
picking out the residue at $u=0$. Using 
$z(2\omega_1x\pm\omega_2)=\lambda(x)\pm{i\over 2}$, and
Eq.~(\ref{defU}) and the heat equation for the Jacobi theta functions
\eqref{heat} we find that
\EQ{S=-{1\over 2i\pi}{d\over d\ttau}\int_{-1/2}^{1/2}
V\left({\lambda(x)}\right)dx.}  
with $\lambda(x)$ given by Eq.~(\ref{eigen})
Since the proposal of \cite{Dijkgraaf:2002dh} implies the
identification $\ttau=\tau/N$ and $2\pi iS=-dW_{\rm eff}/d\tau$ so
that $S$ is the gluino condensate, we must
have
\EQ{W_{\rm eff}=N\int_{-1/2}^{1/2} V\left({\lambda(x)}\right)dx;\qquad
\lambda(x) =-{1\over2}{\theta_3^\prime({\pi x}|\ttau) 
\over\theta_3({\pi x}|\ttau)}.\label{direct}
}
This proves that the eigenvalues of the adjoint scalar 
$\Phi$ in the confining vacuum of the field theory at large $N$ 
are given by the function $\lambda(x)$ and are uniformly
distributed over the interval $x\in[-\tfrac12,\tfrac12]$ {\it
independent of the deformation $V(\Phi)$\/}. Note that, on the contrary
the distribution of eigenvalues of the matrix model depends on the
deformation in a complicated way. 

We should also emphasize that the result \eqref{direct} is also,
according to Dijkgraaf and Vafa, true at finite $N$.
At finite $N$ the 
eigenvalues of $\Phi$ in the confining vacuum have 
been computed from a completely different approach
using the connection between $\N=1^*$ theories and integrable systems
in \cite{DS}. The eigenvalues found in \cite{DS} depend
on a discrete index $j$ and were given as
\EQ{\tilde\lambda_j=-{1\over 2} {\theta_3^\prime(\pi/2-\pi j/N|\tau/N)
\over\theta_3(\pi/2-\pi j/N|\tau/N)};\qquad j=1,\ldots,N\
.\label{inte}
}
(Notice that these, as explained above, are not to be
identified with the eigenvalues $\lambda_j$ of the matrix model!)
The Dijkgraaf-Vafa prediction from the matrix model 
above in Eq.~(\ref{direct}) agrees
precisely with the results of \cite{DS} upon taking the large-$N$
limit where the discrete label $j$ in Eq.\eqref{inte} is replaced by the
continuous 
variable $x$. 
In particular the eigenvalues are uniformly distributed in
$x$. This demonstrates that using the recipe of
\cite{Dijkgraaf:2002fc, Dijkgraaf:2002vw, talk}
the matrix model indeed computes the superpotentials for the
entire class of deformations considered above, provided $N$ is
large. Alternatively, for  
any given deformation it also computes all the condensates
$\VEV{\Tr\,\Phi^n}$ via
\EQ{
\VEV{\Tr\,\Phi^n}=N\int_{-1/2}^{1/2}\, \lambda^n(x)
\,dx\ .\label{cond}
}
In concert with all our previous warning $\VEV{\Tr\,\Phi^n}$, the
condensate of the field theory is {\it not\/} to be identified with the
expectation values in the matrix model \eqref{exp}:
\EQ{
\langle\langle{\rm Tr}\,\Phi^n\rangle\rangle\neq\VEV{\Tr\,\Phi^n}\ .
}

As we have already mentioned, 
the eigenvalues and hence all the condensates in the
confining vacuum are completely 
independent of the actual choice of deformations. This is in complete
accord with expectations from field theory. The 
deformations of the $\N=4$ theory
that we have discussed may be thought of as $\N=1$ perturbations of the
$\N=2^*$ theory. The confining vacua of the resulting theories arise
from points on the Coulomb branch where the Seiberg-Witten curve for
the $\N=2^*$ theory degenerates maximally into a torus of complex
structure $\tau/N$. The location of such a point in moduli space
given by condensates $\VEV{\Tr\,\Phi^n}$, is completely independent of
the choice of $V(\Phi)$. This is in fact expected to be true in all
massive vacua of the $\N=1$ theory.

Although we have seen a remarkable agreement between the
matrix model recipe and the field theory results of
\cite{DS}, strictly this agreement appears when the   
eigenvalues form a continuous distribution and can happen only 
in the large-$N$ limit of the field theory. Indeed, the
appearance of integrals over a continuous label rather discrete sums
in Eq.~(\ref{direct}) suggests that we are really doing a large-$N$
calculation from the point of view of the field theory. So does the
large-$N$ matrix model, or 
the planar graphs of the matrix model really
compute the holomorphic sector of the finite-$N$ field theory? The
resolution of this puzzle will be the subject of Section 3. 

\subsection{Some explicit examples}

We have computed the superpotentials from the matrix model by
explicitly obtaining $G(z(u))$ for $V(\Phi)=\Phi^2$ and $\Phi^4$
and computing $S$, $\partial{\cal F}_0/\partial S$ as
the integrals of $G(z)dz$ over the $A$ and $B$ cycles. These results
match with a direct computation of the superpotential using the
eigenvalue distribution following Eq.~(\ref{direct}). We quote the
results for these cases below

(i) $V(\Phi)=\Phi^2$:
\AL{G(z(u))&=\frac{\omega_1^2}{\pi^2}
\wp(u)-\frac{2w_1\zeta(\omega_1)}{\pi^2}\
,\\
\Pi_A&={dh(\ttau)\over d\ttau}; \qquad\Pi_B =\ttau{dh(\ttau)\over
d\ttau}-h(\ttau).\\
W_{\rm eff}&=-Nh(\tau/N)=-{N\over12}E_2({\tau/N}).}
The additive normalization of $W_{\rm eff}$ can be fixed
by a direct computation of 
Eq.~(\ref{direct}) which gives
\EQ{W_{\rm eff}=N\int_{-1/2}^{1/2}\lambda^2(x)\,dx=-{N\over12}
\big(E_2({\tau/N})-1\big).
\label{W2}
}

(ii) $V(\Phi)=\Phi^4$:
\AL{G(z(u))&={\omega_1^4\over\pi^4}\wp(u)^2+{\omega_1^2\over2\pi^4}\left(\pi^2-
8\omega_1\zeta(\omega_1)\right)\wp(u)\notag\\
&\qquad\qquad\qquad-{\omega_1\over6\pi^4}
\left(g_2\omega_1^3+6\pi^2\zeta(\omega_1)-36\omega_1\zeta(\omega_1)^2\right)\
,\\
\Pi_A&={dh(\ttau)\over d\ttau}; \qquad\Pi_B =\ttau{dh(\ttau)\over
d\ttau}-h(\ttau).\\
W_{\rm eff}&=-Nh(\tau/N)={N\over720}(10E_2({\tau/N})^2-E_4(\tau/N)-30E_2(\tau/N)).}
The additive normalization of $W_{\rm eff}$ can be fixed
by a direct computation of 
Eq.~(\ref{direct}) which gives
\EQ{W_{\rm eff}=N\int_{-1/2}^{1/2}\lambda^4(x)\,dx={N\over720}\big(
10E_2({\tau/N})^2-E_4(\tau/N)-30E_2(\tau/N)+21\big).\label{W4}} 


\subsection{Summary}

In summary, in this section we have verified that the Dijkgraaf-Vafa
prescription for computing $\N=1$ superpotentials from matrix models
works for a large class of deformations of the $\N=4$ theory. In
particular it appears to provide a recipe for extracting quantum field
theory eigenvalue distributions from the matrix model which therefore
permits an exact computation of all higher condensates and
superpotentials for general deformations of the type discussed
above. Our check of this proposal rests on the precise
agreement between these field theory
eigenvalues extracted from the matrix model, and those
computed independently from a completely different approach in
\cite{DS}. 

It must be emphasized that the field theory eigenvalue
distributions are {\it not} identical to the matrix model
distributions. The spectral densities of the two systems are different
although there exists a recipe for extracting the QFT distributions
from those of the matrix model.  

\section{Large $N$ versus finite $N$}

As we have mentioned, one remarkable consequence of the Dijkgraaf-Vafa
hypothesis is that the exact superpotential is captured by a large-$N$
analysis. This presents a real puzzle: we have already remarked that
the matrix model yields the distribution of eigenvalues of $\Phi$
which agrees with the large-$N$ limit of the distribution computed from the
integrable system approach in \cite{DS}; however, surely when we
calculate the condensates $\langle{\rm Tr}\,\Phi^n\rangle$ using the
finite-$N$ distribution of \cite{DS}, we will find differences from
the large-$N$ result.
   
Put another way, the large-$N$ matrix model 
(DV) result for the condensates \eqref{cond} are given
by an integral along one of the cycles on the torus
which can be written in terms of an integral over theta
functions using Eqs.~\eqref{eigen} and \eqref{cond}:
\EQ{
\langle{\rm Tr}\,\Phi^n\rangle_{\text{DV}}
=N\int_{-1/2}^{1/2} 
\left(-\frac12\frac{\theta_3'(\pi x|\tau/N)}{\theta_3(\pi
x|\tau/N)}\right)^n\ dx\ ,
\label{INT}
}
while, in the same normalization, the condensates at finite-$N$ 
can be calculated from the exact eigenvalues that follow from the
integrable system. In the confining vacuum these 
are given in Eq.~\eqref{inte}.
Hence, the condensate which follows from the integrable system
analysis is
\EQ{
\langle{\rm Tr}\,\Phi^n\rangle_{\text{int}}=
\sum_{j=1}^N
\left(-\frac12\frac{\theta_3'(\pi/2-\pi j/N|\tau/N)}{\theta_3(\pi/2-\pi
j/N|\tau/N)}\right)^n\ .
\label{SUM}
}
Clearly, \eqref{INT} and \eqref{SUM} differ at finite $N$ and so, one
might think, that at finite $N$ the 
whole Dijkgraaf-Vafa hypothesis is called into question. 

However, it is well-known that the condensates of the $\N=1^*$ theory
are plagued by mixing ambiguities \cite{oferandus, ambig}. There is no
canonical 
definition of
the operator $\Phi^n$ and different approaches can lead to expressions
which differ by admixtures of $\Phi^p$, $p<n$ (including the identity
operator $p=0$). What is important, however, is that the mixing
coefficients should be vacuum independent. This means that the mixing
coefficients should only depend on modular functions involving the
torus whose complex structure is the bare coupling $\tau$ rather than
the effective coupling in the confining vacuum $\tau/N$. It simply
would not make sense if the mixing coefficients between two approaches
depended on the vacuum---in the present case the confining one. Having
appreciated the possibility of mixing we can now see how the
Dijkgraaf-Vafa hypothesis can be correct even though \eqref{INT} and
\eqref{SUM} are manifestly different at finite $N$. The resolution is
that definitions of $\langle{\rm Tr}\,\Phi^n\rangle_{\text{DV}}$ and
$\langle{\rm Tr}\,\Phi^n\rangle_{\text{int}}$ differ by vacuum
independent mixings.

Although we have no general proof of this, we can present very strong
arguments for its verisimilitude. Firstly consider $n=2$. In this case, the sum
\eqref{SUM} gives
\EQ{
\langle{\rm Tr}\,\Phi^2\rangle_{\text{int}}=-\frac {N}{12}
\big(E_2(\tau/N)+(N-2)E_2(\tau)-(N-1)E_2(\tau/2)\big)\ .
\label{sumn}
}
Notice that \eqref{sumn} actually vanishes for $N=2$ which
follow from the fact that the two eigenvalues vanish in this case.
Notice also that it differs from \eqref{W2} by terms which involve modular
functions of $\tau$ and $\tau/2$, rather than $\tau/N$ 
and, therefore, one suspects that they differ by a
vacuum independent constant which corresponds to mixing with the identity
operator. 

At this point we should remind the reader that the class of theories we are
considering have a large number of vacua with a mass gap, where the
theory is realised in different phases. The canonical examples are the
Higgs vacuum where the gauge group is broken completely by the Higgs
mechanism, and the confining vacuum where classically the full gauge
symmetry is unbroken although quantum mechanically there are no
massless states due to strong coupling effects which cause the
elementary degrees of freedom to confine and generate a mass gap. It
is well-known that the Higgs and confining vacua are actually related
by an $S$-duality, $\tau\rightarrow-1/\tau$, while more general
modular transformations map to other massive vacua \cite{vw, nick,
nickprem}. This provides us with a natural definition for the condensates. 

One way to disentangle the operator ambiguities is to note
there exists a natural definition of the condensates, which we
denote simply as $\langle\Tr\,\Phi^n\rangle$, that have ``good'' modular
properties. What we mean by this is that the value of the condensate
in the Higgs vacuum is given by taking the value in the confining
vacuum with the $S$-duality replacement $\tau\to-1/\tau$ 
and then by multiplying by the factor $\tau^{-n}$
\cite{nick,oferandus}. 
The values in other vacua are also related by
other modular transformations. That such a basis of operators with
good modular properties should exist follows ultimately from the fact
that we are considering a deformation of the $\N=4$ theory, which is
self-dual under modular transformations, by
operators which can be chosen to have a definite weight. By assigning
a compensating modular weight to the couplings the resulting deformed
theory is then modular invariant. However, the vacua of the theory
spontaneously break the modular symmetry. The modular symmetry then
simply permute the vacua. We can also establish the existence of the
modular covariant basis of operators directly from the integrable
system: this tells us how $\langle\Tr\,\Phi^n\rangle$ is related to 
$\langle\Tr\,\Phi^n\rangle_{\rm int}$ and therefore, using \eqref{sumn}
and its generalization to arbitrary deformation, how
$\langle\Tr\,\Phi^n\rangle$ is related to $\langle\Tr\,
\Phi^n\rangle_{\rm DV}$.  

For the $\Phi^2$ operator, the modular covariant
definition of the condensate is related to the matrix one by the
vacuum independent---so involving modular functions of $\tau$ rather
than $\tau/N$---shift
\EQ{
\langle\Tr\,\Phi^2\rangle=\langle\Tr\,\Phi^2\rangle_{\rm DV}
+\frac{N}{12}\big(NE_2(\tau)-1\big)\ .
\label{rel}
}
This then  yields the modular covariant value of the condensate
derived in \cite{nick} in the confining vacuum:
\EQ{
\text{Confining:}\qquad\qquad\langle\Tr\,\Phi^2\rangle=-
\frac{N}{12}\big(E_2(\tau/N)-NE_2(\tau)\big)\ .
\label{conf}
}
The expression for the condensate in the Higgs
vacuum derived in \cite{nick} is
\EQ{
\text{Higgs:}\qquad\qquad\langle\Tr\,\Phi^2\rangle=-
\frac{N^2}{12}\big(NE_2(\tau N)-E_2(\tau)\big)\ .
\label{higgs}
}
As explained above, 
the expressions \eqref{conf} and \eqref{higgs} are related by
\EQ{
\langle\Tr\,\Phi^2\rangle\Big|_{\rm
Higgs}=\tau^{-2}\Big(\langle\Tr\,\Phi^2\rangle\Big|_{\rm
Conf}\Big)_{\tau\to-1/\tau}\ .
}
In order to see this one uses the modular transformation 
property of the Eisenstein series \eqref{mode}.

Now consider the $\Phi^4$ deformation. 
In this case the sum \eqref{SUM} gives
\SP{
\langle{\rm Tr}\,\Phi^4\rangle_{\text{int}}=&\frac {N}{720}
\big(10E_2(\tau/N)^2-E_4(\tau/N)\big)\\
&+\frac16\big((2N-3)E_2(\tau/2)-(2N-6)
E_2(\tau)\big)\langle{\rm Tr}\,\Phi^2\rangle_{\text{int}}
\\
&+\frac{N}{720}\big(5N(N-1)E_2(\tau/2)^2-10N(N-2)E_2(\tau)^2\\
&-(N-1)(N-3)^2E_4(\tau/2)+N^2(N-2)E_4(\tau)\big)\ .
\label{sumn4}
}
In the large-$N$ limit one can easily see that \eqref{sumn4} goes over
to \eqref{W4}. In addition it vanishes for $N=2$ as remarked
above. Notice that the difference between $\langle{\rm Tr}\,
\Phi^4\rangle_{\text{DV}}$ and $\langle{\rm
Tr}\,\Phi^4\rangle_{\text{int}}$ is again by a vacuum
independent, {\it i.e.\/}~$\tau/N$-independent, 
admixture of ${\rm Tr}\,\Phi^2$ and the identity operator.
Once again there is a natural modular covariant definition of the
condensate which can be established by investigating the integrable
system and establishing a relation between  $\langle{\rm
Tr}\,\Phi^4\rangle$ and $\langle{\rm
Tr}\,\Phi^4\rangle_{\text{int}}$. Using \eqref{sumn4} one eventually
finds 
\SP{
\langle\Tr\,\Phi^4\rangle&=\langle\Tr\,\Phi^4\rangle_{\rm DV}+
\frac1{12}\big(6-NE_2(\tau))\langle\Tr\,\Phi^2\rangle_{\rm DV}\\
&\qquad\qquad+\frac N{720}\big(N^2E_4(\tau)-
5N^2E_2(\tau)^2-35NE_2(\tau)+9\big)\ .
}
The result \eqref{W4}, then gives the value of the condensate in the
confining vacuum:
\SP{
\text{Confining:}\qquad\qquad\langle\Tr\,\Phi^4\rangle
&=\frac{N}{720}\big(10E_2(\tau/N)^2-E_4(\tau/N)\\
&-5N^2E_2(\tau)^2+N^2E_4(\tau)-5
NE_2(\tau)E_2(\tau/N)\big)\ .
}
The value in the Higgs vacuum follows by taking $\tau\to-1/\tau$ and 
multiplication by $\tau^{-4}$ yielding
\SP{
\text{Higgs:}\qquad\qquad\langle\Tr\,\Phi^4\rangle
&=\frac{N}{720}\big(10N^4E_2(\tau N)^2-N^4E_4(\tau N)\\
&-5N^2E_2(\tau)^2+N^2E_4(\tau)-5
N^3E_2(\tau)E_2(\tau N)\big)\ . 
}

\section{Perturbation of the Leigh-Strassler deformation}

We now turn to an application of the techniques of Dijkgraaf and Vafa
to a very different class of deformations of the $U(N)$ $\N=4$
theory. 
We consider the $\N=4$ theory perturbed both by a certain
exactly marginal operator (the Leigh-Strassler deformation \cite{LS})
and certain mass terms. (The authors of \cite{Dijkgraaf:2002dh} have
pointed out that such deformations  
can accessed via known matrix models.)
The Leigh-Strassler deformations of the $\N=4$
theory lead to a 3-dimensional fixed manifold of $\N=1$
theories, via special trilinear perturbations of the $\N=4$
superpotential. Among these marginal deformations, of particular
interest is the so-called ``$q$-deformation'' of the $\N=4$
superpotential,
\EQ{W = \Tr\big(i\Phi[\Phi^+,\Phi^-]_\beta\big) \equiv
\Tr\big(i\Phi[\Phi^+\Phi^-e^{i\beta/2}-\Phi^-\Phi^+e^{-i\beta/2}]\big).}

We consider a special perturbation of this theory via the
superpotential 
\EQ{W=\Tr\big(i\Phi[\Phi^+,\Phi^-]_\beta+\Phi^+\Phi^-
+\Phi^2\big)\label{LSpert}} 
We choose this particular form for the mass perturbation as the
corresponding large-$N$ matrix model can be solved \cite{kostov}
around the trivial classical solution $\Phi=\Phi^\pm=0$. This trivial
solution corresponds of course to the confining vacuum of the
$\N=1$ gauge theory.

As demonstrated in \cite{kostov}, expanding around the trivial vacuum
the fields $\Phi^\pm$ can be 
integrated out and the corresponding one-matrix model can be solved
in the large-$N$ limit by going to an eigenvalue basis and a change of
variables. We simply quote this change of variables here, without
further explanation. In particular the eigenvalues $\lambda_i$ of
$\Phi$ are redefined in terms of new variables $\delta_i$
\EQ{\lambda_i=-e^{\delta_i}+{1\over2{\rm sin}(\beta/2)}\label{redef}} 
The eigenvalues experience a force that causes them to
condense and expand out into a cut in the large-$N$ limit. The complex
$z$-plane in which the $\delta_i$'s live turns out to be a cylinder
with two cuts and gluing conditions on the two cuts implied by the
saddle point equation for the function $G(z)$. 

\subsection{The $\ttau$ torus}

The key point for us is that large-$N$ matrix model solution once
again leads to the emergence of a torus. 
The $z$-plane being a two-cut cylinder with certain gluing conditions,
there is a natural map to a torus $E_\ttau$ with 2
marked points. As before, we think of this torus as the complex $u$-plane modded
out by lattice translations. The map $z(u)$ and $G(z(u))$ can be
determined precisely using in particular the asymptotic behaviour of
$G(z(u))$. We refer the reader to \cite{kostov} for the details of
these asymptotic properties. Using these properties we are able to
determine $G(z(u))$ and the map $z(u)$ in the elliptic parametrization as
\EQ{G(z(u))={\omega_1^2\over\pi^2}\;{{\rm cos}(\beta/2)\over 2{\rm
sin}^3(\beta/2)}\;{\theta_1(\beta/2)^2\over 
\theta_1^\prime(\beta/2)^2}\;
\big[{\wp(u+\omega_1\beta/2\pi)-\wp(\omega_1\beta/\pi)}\big]}
and
\EQ{\exp{z(u)}={1\over 2}{\rm cot}(\beta/2)\;
{\theta_1^\prime(0)\over\theta_1^\prime(\tfrac\beta
2)}{\theta_1(\pi u/2\omega_1-\beta/4)\over\theta_1(\pi u/2\omega_1+\beta/4)}.}
Note that $G(z(u))$ is elliptic while $z(u)$ is periodic only along
the $A$-cycle while $B$-cycle shifts lead to shifts in $z(u)$,
\EQ{z(u+2\omega_1)=z(u);\qquad z(u+2\omega_2)=z(u)+i\beta.}
This encodes the gluing conditions implied by the saddle point equation
satisfied by
$G(z)$. Note that the points $u=\pm\beta\omega_1/2\pi$ are the two marked points
on the $\ttau$-torus and they map to the points $z=\mp \infty$. Note
also that in the $\beta\rightarrow 0$ limit $z(u)$ and $G(z(u))$ reduce to
Eqs.(\ref{defz}) and \eqref{gquad} after a simple rescaling of $z(u)$
by $\beta$.

\subsection{The effective superpotential}

We can now use the Dijkgraaf-Vafa prescription to compute the
effective superpotential for the relevant deformation of the Leigh
Strassler theory. We have to compute the integrals of the one-form
$G(z)dz$ over the $A$ and $B$ cycles of the $\ttau$-torus. The
integrals are surprisingly simple and we find
\EQ{\Pi_A=-i\int_A G(z(u)) z^\prime(u) du ={d h(\ttau)\over
d\ttau};\quad \Pi_B=-i\int_BG(z(u)) z^\prime(u) du =\ttau{d h(\ttau)\over
d\ttau}-h(\ttau)}
where
\EQ{h(\ttau)={{\rm cos}(\beta/2)\over 4 {\sin}^3(\beta/2)}\;\;
{\theta_1(\beta/2|\ttau)\over\theta_1^\prime(\beta/2|\ttau)}.}
Thus we find the same structure as before for $S$ and $\partial {\cal
F}_0/\partial S$ which guarantees that 
\EQ{dW_{\rm eff}/dS=0\ \Longrightarrow\  \ttau=\tau/N}
and 
\EQ{W_{\rm eff}=-Nh=-N{{\rm cos}(\beta/2)\over 4 {\sin}^3(\beta/2)}\;\;
{\theta_1(\beta/2|\tau/N)\over\theta_1^\prime(\beta/2|\tau/N)}}
up to an additive constant. A quick check confirms that as
$\beta\rightarrow 0$ this result reproduces the superpotential for
$\N=1^*$ theory.\footnote{ Actually there is a $\tau$-independent
additive constant which diverges 
in the $\beta\rightarrow 0$ limit which is a consequence of the linear shift
of variables introduced in \cite{kostov}. Strictly speaking, the resultant shift
in the
matrix model action $N/(4{\sin}^2(\beta/2))$ needs to be reintroduced into
$W_{\rm eff}$ which
would then get rid of the offending additive term in the
$\beta\rightarrow 0$ limit.}

\subsection{QFT eigenvalues and condensates}

We recall that in the deformations considered in Section 2, the map
from the $A$-cycle of the the torus $E_\ttau$ to the $z$-plane turned
out to yield the field theory eigenvalues distributed uniformly on
this cycle of $E_\ttau$. We can ask if a similar picture emerges for
perturbations
of the Leigh-Strassler theory. We first note that we need to take into
account the field redefinition Eq.\eqref{redef} and so we need to
consider the function $-\exp z(u)+1/(2{\rm sin}(\beta/2))$ evaluated
along the $A$-cycle of the torus, and we find that the relevant
function (after a shift of $z$ by $i\beta/2$) is
\EQ{\lambda(x)=-{1\over 2}{\rm cot} (\tfrac \beta 2)\;
{\theta_1^\prime(0)\over\theta_1^\prime(\beta/
2)}
 {\theta_3(\pi x -\beta/4)\over\theta_3(\pi x +\beta/4)}+{1\over 2{\rm
sin}(\beta/2)};\qquad x\in[-\tfrac12,\tfrac12].\label{LSeigen}}

It is quite remarkable that for the quadratic deformation considered
in this section we find that indeed,
\EQ{W_{\rm eff}= N\int_{-1/2}^{1/2} \lambda^2(x)dx =-N{{\rm
cos}(\beta/2)\over 4
{\sin}^3(\beta/2)}\;\;{\theta_1(\beta/2|\tau/N)\over\theta_1^\prime(\beta/2|\tau/N)}+ 
{N\over 4{\rm sin}^2(\beta/2)}}  
Although we don't have a general proof, the above leads us to believe
that $\lambda(x)$ indeed yields the eigenvalues of the adjoint scalar
in the confining vacuum of the field theory, distributed uniformly
over the interval $[-{1\over 2},{1\over 2}]$. In other words, replacing
$x$ by a discrete index as in the previous section would precisely yield
the eigenvalues of the adjoint scalar in the confining vacuum of the
finite $N$ field theory. 
We further conjecture that our conclusions in
Section 3 on the relation between the field theory quantities at
large-$N$ and finite $N$ should also hold for the deformations of the
Leigh-Strassler theories. Note also that our earlier comments on the
distinction between field theory eigenvalue densities and matrix model
eigenvalue densities apply in this case as well.

{\it A potential test involving $S$-duality}:

Although not completely understood in the context of this class of
theories, one expects the Higgs and confining phases of these theories
to be related by $S$-duality {i.e.} $\tau\rightarrow -1/\tau$.
A possible non-trivial test of our results for the superpotential and 
eigenvalue distributions in the confining vacuum is  a comparison with
classical eigenvalues in the Higgs vacuum after performing $S$-duality
on the confining vacuum and taking the semiclassical limit. It is not
difficult to see that the
classical eigenvalues in the Higgs vacuum are distributed on a circle
with $\tilde\lambda_j\sim \exp(-i(N+1-2j)\beta/2)/{\rm cos}(N\beta/2)$; $j=1,\ldots N$. After an
$S$-duality transformation, and redefining 
$\beta^\prime=\tau\beta$, in the semiclassical $\tau\rightarrow
i\infty$ limit
our result Eq.\eqref{LSeigen} also yields a uniform circular
distribution of eigenvalues $\lambda(x)\sim\exp(-i\beta^\prime N
x)/{\rm cos}(\beta N/2)$; $x\in[-1/2,1/2]$. This gives a hint of how $S$-duality should act on the
Leigh-Strassler deformation parameter. Indeed the Leigh-Strassler
family of $\N=1$ CFTs are not self-dual under $SL(2,{\mathbb Z})$ \cite{LS}. Our
observations here could provide a starting point for further
investigation into these issues.

\section{Conclusions}

In this article we have performed a test of the proposal of
\cite{Dijkgraaf:2002fc,Dijkgraaf:2002vw,talk} for a large class of
$\N=1$ preserving deformations of the $\N=4$ theories in their
confining phase. We have shown
that the Dijkgraaf-Vafa proposal translates into a rather simple
recipe for extracting eigenvalues of the adjoint scalar $\Phi$ in the
confining vacuum from the corresponding
matrix model, thereby yielding all the relevant condensates including
the superpotential in that vacuum. We find that the eigenvalues
extracted using this recipe match precisely with the results of
\cite{DS} wherein the relevant eigenvalues were obtained from a
completely different approach involving the connection of $\N=1^*$
theories with integrable models. Although at first sight the matrix
model recipe yields large-$N$ field theory results, we demonstrate through
certain
nontrivial examples that these large-$N$ results are simply related to
their finite $N$ counterparts via vacuum independent field
redefinitions or operator mixings. Thus the large-$N$ matrix model
indeed computes the holomorphic observables of the field theory at
finite $N$ in its confining vacuum.

We also compute the effective superpotential in the confining vacuum
of a mass deformation of the Leigh-Strassler CFT. We demonstrate that
for this class of deformations, the matrix model once again yield the
field theory eigenvalues and condensates via a simple recipe. A
potential test of our results in the confining vacuum involves
utilising $S$-duality to obtain the eigenvalues in the Higgs vacuum in
the classical limit. The latter can be computed independently and we find
nontrivial agreement if we assume that the Leigh-Strassler deformation
parameter transforms with modular weight one, under S-duality.

An interesting question concerns whether the matrix model approach can capture
the full vacuum structure of the $\N=1^*$ theory, including the Higgs
vacuum. We answer this in the affirmative in a separate publication.

{\bf Acknowledgements} We would like to thank R. Dijkgraaf and C. Vafa
for extensive stimulating discussions. S.P.K. would like to
acknowledge support from a PPARC Advanced Fellowship. 

\startappendix

\Appendix{Some Properties of Elliptic Functions}

In this appendix we provide some useful---but far from
complete---details of elliptic functions and their near cousins.
For definitions and a more complete
treatment, we refer the reader to one of the textbooks, for example
\cite{WW}. An elliptic function $f(u)$ is a function on the complex
plane, periodic in two periods $2\omega_1$ and $2\omega_2$. We will
define the lattice $\Gamma=2\omega_1{\mathbb Z}\oplus2\omega_2{\mathbb
Z}$ and define the basic period parallelogram as
\EQ{
{\cal
D}=\big\{u=2\mu\omega_1+2\nu\omega_2,\ 0\leq\mu<1,\ 0\leq\nu<1\big\}\ .
}
The complex structure of the torus defined by identifying the edges of
${\cal D}$ is
\EQ{
\tau=\omega_2/\omega_1
}
and we also define
\EQ{
q=e^{i\pi\tau}\ .
}

\subsection{The Weierstrass function}

The archetypal
elliptic function is the Weierstrass $\wp(u)$ function. It is an even
function which is 
analytic throughout ${\cal D}$, except at $u=0$
where it has a double pole:
\SP{
&\wp(u)=\frac1{u^2}+\sum_{k=1}^\infty c_{k+1}u^{2k}\ ,\\
&c_2=\frac{g_2}{20}\ ,\quad c_3=\frac{g_3}{28}\ ,\quad
c_k=\frac{3}{(2k+1)(k-3)}\sum_{j=2}^{k-2}c_jc_{k-j}\quad k\geq4\ .
}
The Weierstrass function satisfies the fundamental identity
\EQ{
\Big(\frac{d}{du}\wp(u)\Big)^2=4\wp(u)^3-g_2\wp(u)-g_3
}
which defines the {\it Weierstrass invariants\/} 
$g_{2,3}=g_{2,3}(\omega_\ell)$ associated
to the torus.

\subsection{The Weierstrass zeta function}

We are also interested in other functions which are only
quasi-elliptic. First we have 
$\zeta(u)$. It is an odd function 
with the quasi-elliptic property:
\EQ{
\zeta(u+2\omega_\ell)=\zeta(u)+2\zeta(\omega_\ell)\ .
}
Its derivative gives minus the Weierstrass function
\EQ{
\qquad\wp(u)=-\zeta'(u)\ .
}
It follows that 
$\zeta(u)$ has a simple pole at $u=0$.
Useful identities are
\AL{
&\omega_2\zeta(\omega_1)-\omega_1\zeta(\omega_2)=\frac{\pi i}2\ ,\\
&\zeta(\omega_1+\omega_2)=\zeta(\omega_1)+\zeta(\omega_2)\ ,\\
&\wp(\omega_1+\omega_2)+\wp(\omega_1)+\wp(\omega_2)=0\ .
}

\subsection{The Jacobi theta functions}

We are also interested in the Theta functions $\theta_i(x|\tau)$, or
$\theta_i(x,q)$, $i=1,2,3,4$. They are also quasi-elliptic
functions on ${\cal D}$ when $x=\pi u/2\omega_1$. Each of them satisfies
the heat equation
\EQ{
\pi i\frac{\partial^2\theta_i(x|\tau)}{\partial x^2}+
4\frac{\partial\theta_i(x|\tau)}{\partial\tau}=0\ .
\label{heat}
}
They are related to the
previous functions; for instance, two relations that we need are
\EQ{
\zeta(u)-\frac{\zeta(\omega_1)}{\omega_1}u=\frac{\pi}{2\omega_1}
\left.\frac{\theta'_1(x|\tau)}{\theta_1(x|\tau)}\right|_{x=\pi u/2\omega_1}\equiv
-\frac{i\pi}{2\omega_1}+\frac\pi{2\omega_1}
\left.\frac{\theta'_3(x|\tau)}{\theta_3(x|\tau)}\right|_{x=\pi(u-\omega_1-\omega_2)/2w_1}
\ ,
\label{relt}
}
where the derivative is with respect to $x$. 

\subsection{The Eisenstein series}

Finally, we introduce the Eisenstein series and their relation to
elliptic functions. There are a number of ways to define these series:
see \cite{Kob}\footnote{Beware, the 
definition of $q$ in that reference is the square
of ours.}; for instance
\EQ{
E_k(\tau)=\frac12\sum_{m,n\in{\mathbb
Z}\atop(m,n)=1}\frac1{(m\tau+n)^k}\ ,
}
where $(m,n)$ denotes the greatest common divisor. Each series has a
$q$-expansion; for example
\SP{
E_2(\tau)&=1-24\sum_{n=1}\sigma_1(n)q^{2n}\ ,\\
E_4(\tau)&=1+240\sum_{n=1}\sigma_3(n)q^{2n}\ ,\\
E_6(\tau)&=1-504\sum_{n=1}\sigma_5(n)q^{2n},
}
where $\sigma_j(n)$ is a sum over each positive integral divisor of
$n$ raised to the $j^{\rm th}$ power. 

Useful identities are
\SP{
E'_2&=\frac{\pi i}6\big(E_2^2-E_4\big)\ ,\\
E'_4&=\frac{2\pi i}3\big(E_4E_2-E_6\big)\ ,\\
E'_6&=\pi i\big(E_6E_2-E_4^2\big)\ ,
}
where the derivatives are defined with respect to $\tau$. Under a
$S$-duality transformation we have
\EQ{
E_2(-1/\tau)=\tau^2E_2(\tau)+\frac{6\tau}{\pi i}\ ,\qquad
E_k(-1/\tau)=\tau^{k}E_k(\tau)\quad k\geq4\ .
\label{mode}
}
The relations of the Eisenstein series to the elliptic functions
defined previously appears through the identities
\EQ{
\zeta(w_1)w_1=\frac{\pi^2}{12}E_2(\tau)\ ,\quad
g_2(2w_1)^4=\frac{4\pi^4}3E_4(\tau)\ ,\quad
g_3(2w_1)^6
=\frac{8\pi^8}{27}E_6(\tau)\ .
}
{
\Appendix{Stationary points of the Hamiltonians}
In this Appendix we show that the configuration of the ECM 
system considered in the text is an equilibrium point with respect to
each of the Hamiltonians $H_{p}={\rm Tr} L^{p}$. In each case Hamilton's
equations read $\partial P_{j}/\partial t=-\partial H_{p}/\partial
X_{j}$ and $\partial X_{j}/\partial t=\partial H_{p}/\partial
P_{j}$. Thus we need to show that the right hand side of both these
equations vanishes in the configuration of interest where $X_{j}=\langle
X_{j} \rangle=2\pi i j/N$ for $j=1,2\ldots N$ and $P_{j}=\langle
P_{j}\rangle =0$. (Note that here for simplicity we present the requisite proof for
the equilibrium positions that correspond to the Higgs vacuum of the
gauge theory. The proof for the confining vacuum can be obtained along
the same lines by using the Lax matrix appropriate for that vacuum and
equilibrium positions $X_{j}=\langle
X_{j} \rangle=2\pi i \tau j/N$ for $j=1,2\ldots N$ and $P_{j}=\langle
P_{j}\rangle =0$.)

Importantly we must also remember to impose the
constraint appropriate to the $SU(N)$ theory:
$H_{1}=\sum_{j=1}^{N}P_{j}=0$. This is most easily accomplished by
introducing a Lagrange multiplier $\mu$ and looking for stationary
points of the modified Hamiltonians $\tilde{H}_{p}=H_{p}-\mu H_{1}$. 
Thus we must show that the equations,   
\begin{eqnarray}
\frac{\partial H_{p}}{\partial X_{j}}=0 & \qquad{} \qquad{} & 
\frac{\partial H_{p}}{\partial P_{j}}=\mu \nonumber \\ 
\label{lag}
\end{eqnarray}
Can be solved in the equilibrium configuration. We will first
demonstrate that,  
\be
\left. \frac{\pa H_p}{\pa X_j} \right|_{\langle X_j \rangle,\langle P_{j}\rangle} =0
\ee  
The Hamiltonians of the integrable system can be expressed with
the Lax matrix as
\be
H_p = {\rm Tr}\;[L^p]
\ee
so we would like to prove that for the Calogero-Moser system
\be
\left. \frac{\pa H_p}{\pa X_j} \right|_{\langle X_j \rangle,\langle P_{j}\rangle} = \left.
{\rm Tr} \left[ L^{p-1} \frac{\pa L}{\pa X_j} \right]
\right|_{\langle X_j \rangle,\langle P_{j}\rangle} =0
\ee

The CM system is described with the Lax matrix
\be
L_{jk} = p_j \delta_{jk} + i (1 - \delta_{jk}) x(X_j - X_k)
\ee
where the function $x(u)$ is given explicitly in \cite{DS}. For the
present purposes we only need two properties of $x(u)$. Firstly it has
period $2\pi i$: $x(u+2\pi i)=x(u)$. Secondly it is an odd function:
$x(-u)=-x(u)$. 
As above, the equilibrium positions of the integrable system are
\be
\langle X_j \rangle = \frac{2 i \pi j}{N} - 
\frac{ i \pi (N-1)}{N} \quad j=1 \ldots N
\ee
and $\langle P_{j} \rangle =0$
The key point is that the Lax matrix becomes a {\em circulant} when
evaluated in this configuration. It is then easy to show that  
the unitary transformation which diagonalises the Lax matrix is, 
\bea
L &=& U^{\dagger} \Lambda U \nonumber \\
\Lambda_{jk} &=& \lambda_j \delta_{jk}
\eea
with
\be
U_{jk} = \frac{1}{\sqrt{N}} e^{\frac{2 \pi i}{N} (j-1)(k-1)}
\ee
The trace then can be written
\be
\left. Tr \left[ L^{p-1} \frac{\pa L}{\pa X_j} \right]
\right|_{\langle X_j \rangle,\langle P_{j}\rangle} = \left.
Tr \left[ \Lambda^{p-1} U \frac{\pa L}{\pa X_j} \right|_{\langle X_j \rangle,\langle P_{j}\rangle} 
 U^{\dagger} \right]
\ee

Substituting the expressions for the Lax matrix and for $U$ the explicit form
of the trace is
\be
\left.
Tr \left[ \Lambda^{p-1} U \frac{\pa L}{\pa X_j} \right|_{\langle X_j \rangle,\langle P_{j}\rangle}
 U^{\dagger} \right] = - \frac{2}{N} 
\sum_{k=1}^{N} \lambda_k^{p-1} 
\sum_{l=1, l \neq j}^{N} x' (X_j - X_l) \sin{\frac{2 \pi}{N} (k - 1) (j - l)} 
\ee
Using that $x'(u)$ is an even function and is periodic in $\omega_{1} =
2 \pi i$ we find the terms in the summation pairwise cancel. Thus the 
summation gives zero and we conclude
\be
\left. \frac{\pa H_p}{\pa X_j} \right|_{\langle X_j \rangle,\langle P_{j}\rangle} = 0
\ee
}
Similarly its easy to show that the second equation in (\ref{lag}) is
solved by choosing $\mu= H_{p-1}$. This completes the proof.

\end{document}